\documentclass[aps,prl,twocolumn,superscriptaddress,letterpaper]{revtex4-1}
\pdfoutput=1
\usepackage{amsmath,amsfonts,amssymb,amsbsy,amscd,amsgen}
\usepackage[usenames,dvipsnames]{xcolor}
\usepackage{graphicx}
\newcommand{\reffig}[1]{Fig.~\ref{#1}}
\renewcommand{\vec}[1]{{\bf #1}}

\begin{document}

\title{Increasing lifetimes and the growing saddles of shear flow turbulence}

\author{Tobias Kreilos}
\affiliation{Max Planck Institute for Dynamics and Self-Organization, Am Fassberg 17, 37077 G\"ottingen, Germany}
\affiliation{Fachbereich Physik, Philipps-Universit\"at Marburg, 35032 Marburg, Germany}
\author{Bruno Eckhardt}
\affiliation{Fachbereich Physik, Philipps-Universit\"at Marburg, 35032 Marburg, Germany}
\affiliation{J.M. Burgerscentrum, Delft University of Technology, Mekelweg 2, 2628 CD Delft, The Netherlands}
\author{Tobias M. Schneider}
\affiliation{Max Planck Institute for Dynamics and Self-Organization, Am Fassberg 17, 37077 G\"ottingen, Germany}

\date{\today}

\begin{abstract}
In linearly stable shear flows turbulence spontaneously decays
with a characteristic lifetime that varies with Reynolds number.
The lifetime sharply increases with Reynolds
number so that a possible divergence marking the transition to
sustained turbulence at a critical point has been discussed.
We present a mechanism
by which the lifetimes increase: 
in the system's state space, turbulent motion is supported by a chaotic saddle.
Inside this saddle a locally attracting periodic orbit is created and undergoes a
traditional bifurcation sequence generating chaos. The formed new
'turbulent bubble' is initially an attractor supporting
persistent chaotic dynamics. Soon after its creation
it collides with its own boundary, by which it becomes leaky and dynamically
connected with the surrounding structures.
The   complexity of the chaotic saddle that supports transient turbulence
hence increases by incorporating the remnant of a new bubble. As a a
result, the time it takes for a trajectory to leave
the saddle and decay to the laminar state is increased.
We demonstrate this 
phenomenon in plane Couette flow and show that
characteristic lifetimes vary non-smoothly and non-monotonically with
Reynolds number. 
\end{abstract}



\maketitle

The emergence of turbulence in shear flows such as
pipe or Couette flow has puzzled scientists for more than a
century, because the transition is not associated with a linear instability
of the laminar base flow \cite{Grossmann2000} and requires perturbations
of finite amplitude \cite{Darbyshire1995,Schmiegel1997}.
Once triggered, turbulence
in these flows is not persistent but can spontaneously decay \cite{Eckhardt2007}.
This phenomenology can be understood in the
context of dynamical systems,
where the structure of the state space - the space of all instantaneous velocity 
fields - is investigated.
The state space is filled with a multitude of unstable invariant
solutions and a turbulent trajectory transiently visits those solutions before it
eventually escapes their neighborhood and decays to the laminar state.
This picture of a \emph{chaotic saddle} that supports transient turbulence
has been corroborated by the direct identification of numerous
invariant solutions and some dynamical connections between them both in plane
Couette and pipe flow \cite{Eckhardt2008b,Kreilos2012,Kawahara2012}.
Further support
comes from the distribution of transient lifetimes, which is
found to be an exponential and thus of the form expected
for chaotic saddles.
The characteristic time of the exponential distribution
increases sufficiently rapidly with Reynolds number Re to explain
why decays are rarely observed at high Re.
How exactly the characteristic lifetime $\tau$ increases has attracted much
recent attention, specifically the question whether there is a
critical Re at
which lifetimes diverge and turbulence becomes sustained
\cite{Grebogi1986,Bottin1998,Hof2006,Peixinho2006,Willis2007,Hof2008,Schneider2008a,Schneider2010b,Eckhardt2007,
Kerswell2005,Crutchfield1988,Kadanoff1984,Duguet2008,Gibson2008,Vollmer2009,vanVeen2011}.

While previous studies have characterized a functional form
of the dependence of $\tau$ on Re,
we here present a mechanism that leads to increasing lifetimes.
Within the picture of a turbulent 
trajectory meandering between invariant solutions before eventually
escaping the chaotic saddle and decaying to laminar flow, increasing turbulent
lifetimes are directly linked to changes of the chaotic saddle
structure.
We describe a sequence of events within a small range of Re that add an
infinite number of invariant solutions to the chaotic saddle.
Thereby the complexity of the chaotic saddle grows and lifetimes increase.
During the process the lifetimes diverge for some Reynolds numbers,
leading to a non-monotonic and non-smooth variation of $\tau$ with Re.

In extended flow systems, the local relaminarization of turbulence can be overcome by
spatial spreading so that turbulence may become sustained in a
percolation-like phase transition despite its local decay
\citep{Allhoff2012,Avila2011}. To
separate the strongly increasing variations of $\tau$ in a confined domain from
the spatial spreading of turbulence, we consider plane Couette flow in
a small periodic box. 
The box is chosen such that the flow is described by a comparably low number of degrees
of freedom ($\mathcal{O}(10^5)$).
This allows for a detailed analysis of the high-dimensional 
state-space, which remains computationally challenging;
specifically we consider the
system studied previously in \citet{Kreilos2012}, which is a bit longer and narrower than
the minimal flow unit studied for example in \citep{Waleffe1998}.
Plane Couette flow is the flow of a fluid between two parallel plates moving
at constant speed in opposite direction governed by
the incompressible Navier Stokes equations $\partial_t \vec u + (\vec \cdot u\nabla)\vec u
= -\nabla p + \mathrm{Re}^{-1} \Delta \vec u$
together with the continuity equation $\nabla \cdot \vec u = 0$
and boundary conditions
$\vec u=\pm \vec e_x$ at $y\pm 1$, where p is the pressure, y the wall-normal
coordinate. The Reynolds number is defined by $Re = U_0 h / \nu$ with $U_0$ half the
relative plate velocity, $h$ half the plate separation and $\nu$ the
viscosity of the fluid and using $U_0$ and $h$ as velocity- and length-scale, respectively.
Direct numerical simulations are performed using the
pseudo-spectral solver channelflow \citep{Gibson2008,channelflow} developed by John F. Gibson.
The computational domain is of size $L_x\times
2h\times L_z = 2\pi\times2\times\pi$ with a resolution of 
$48\times33\times48$ modes subject to periodic boundary
conditions in the homogeneous directions and enforced shift-and-reflect
symmetry $(u,v,w)[x,y,z] \equiv \sigma_{SR} (u,v,w)[x,y,z] = (u,v,-w)[x+L_x/2,y,-z]$.

Plane Couette flow is linearly stable for all Re and the first 
(known) non-trivial invariant solutions are the Nagata-Busse-Clever (NBC)
states \cite{Nagata1990,Clever1997,Waleffe1998}, a pair of fixed points that
are created in a saddle-node bifurcation.
In our domain the bifurcation takes place at $Re=163.8$ and 
the upper-branch solution (with higher energy) is attracting while the lower
branch has a single unstable direction.
As a consequence,
its stable manifold forms the separating boundary between
trajectories approaching
the upper branch and the laminar state \cite{Skufca2006,Schneider2008}.
Increasing Re, the upper branch undergoes a period doubling cascade, generating chaos.
At $Re=188.51$ the chaotic attractor formed around the upper branch 
subsequently collides with the 
lower branch. In this boundary crisis \cite{Kreilos2012}, the chaotic attractor 
becomes ``leaky'' and turns into a chaotic saddle, from which
trajectories eventually escape. Similar scenarios
have been found
for periodic \cite{Mellibovsky2012} and
localized \cite{Avila2013} structures in pipe flow
as well as in magnetohydrodynamics \cite{Riols2013},
indicating a generic mechanism by which the saddle that supports
first short-lived turbulent transients in shear flows is created.
Similar scenarios are well-known in low-dimensional dynamical systems
\cite{Grebogi1983,Lai1994,Lai2011}.

\begin{figure}
  \includegraphics[width=\linewidth]{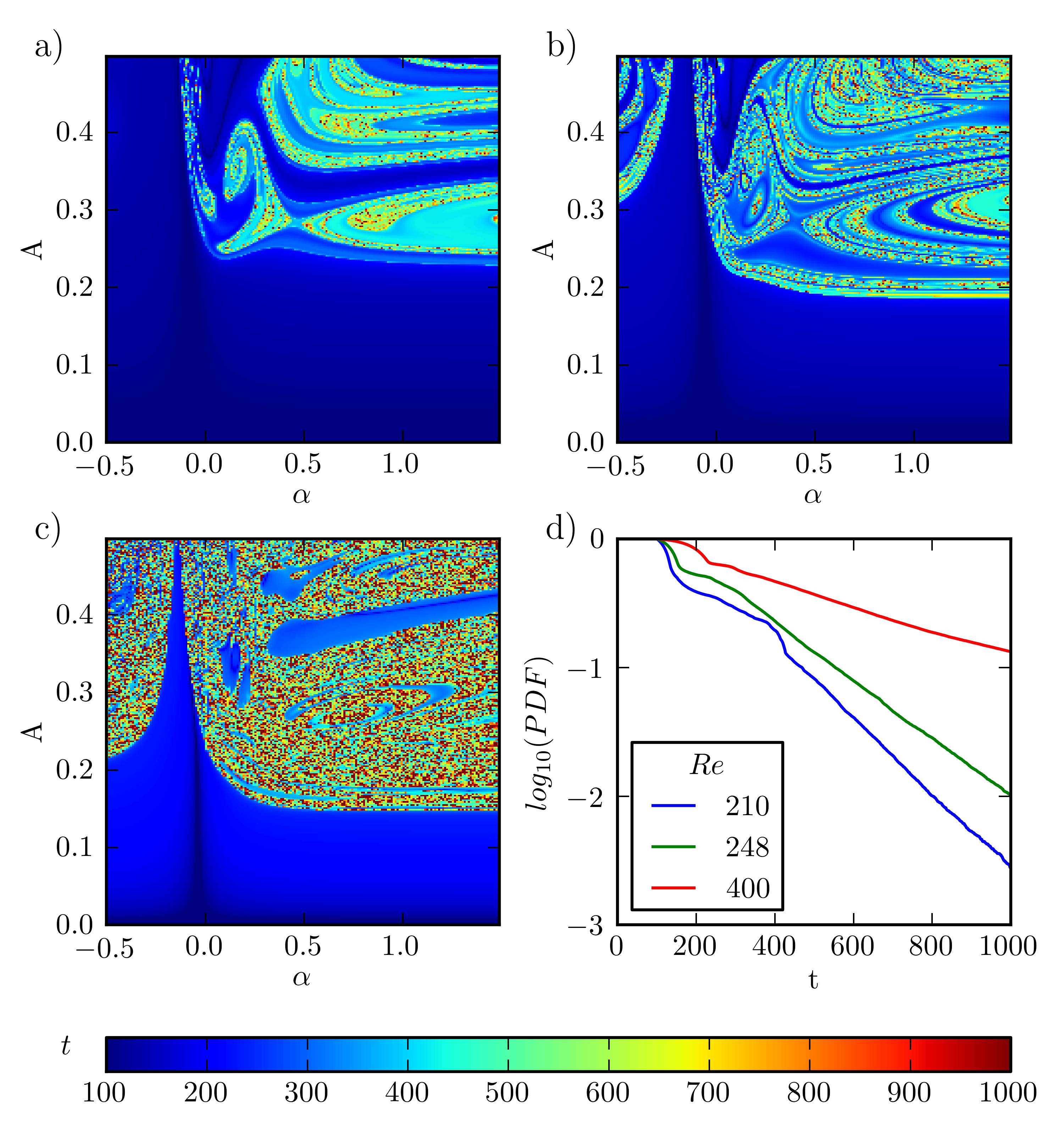}
  \caption{\label{fig:SaddleIntro}
    (color online)
    (a) - (c): The growing turbulent saddle in state space for different Reynolds numbers, visualized
    by the lifetime of initial conditions in a 2D section.
    The 50,000 initial conditions in each panel
    are chosen by interpolating between lower- and upper-branch NBC-solutions
    on the x-axis  and rescaling the perturbation amplitude $A$ on the
    y-axis. The x-axis is rescaled such that the $L_2$-distance of the NBC-states
    is proportional to their distance in the projection, giving a physical meaning to
    the scales in the figure:
    $u = A  u_\alpha / \| u_\alpha \|, u_\alpha = u_{LB} + \alpha(u_{UB}-u_{LB}) / \sqrt{\|u_{UB} - u_{LB}\|^2 - (\|u_{UB}\|-\|u_{LB}\|)^2}$;
    In the sequence a) $Re=210$. b) $Re=248$. c) $Re=400$, the saddle
    grows while it develops ever finer striations, indicating the structures supporting the saddle to get denser.
   d) Survival probabilities for the presented Reynolds numbers show
   clear exponential tails with increasing slope, indicating a characteristic lifetime which
   increases with Re.}
\end{figure}

As Re increases, the chaotic saddle grows to fill a larger part
of state space. In \reffig{fig:SaddleIntro} this is
visualized for three successively increasing Re by plotting the
lifetimes of initial conditions, see figure caption for details.
Blue regions correspond to initial conditions
that directly laminarize while the chaotic saddle appears as colorful fractal regions of rapidly
varying lifetimes indicative of the sensitive dependence on 
initial condition.
Since the lower branch solution has a single unstable direction, its stable
manifold is of co-dimension one and the general topology of state space can
be understood in a two-dimensional section.
For Re=210 (panel \textit{a}),
the saddle only 
occupies a small area. As Re increases (panels \textit{b} and \textit{c}), the colorful fractal
regions marking the saddle both grow and the variations inside become denser.
As expected, the survival probabilities, shown in panel \textit{d}, show exponential tails 
indicating a constant escape probability from the saddle.
Their slope, the characteristic lifetime $\tau$,
increases with Re as the saddle structure fills the domain.

\begin{figure}
  \includegraphics[width=\linewidth]{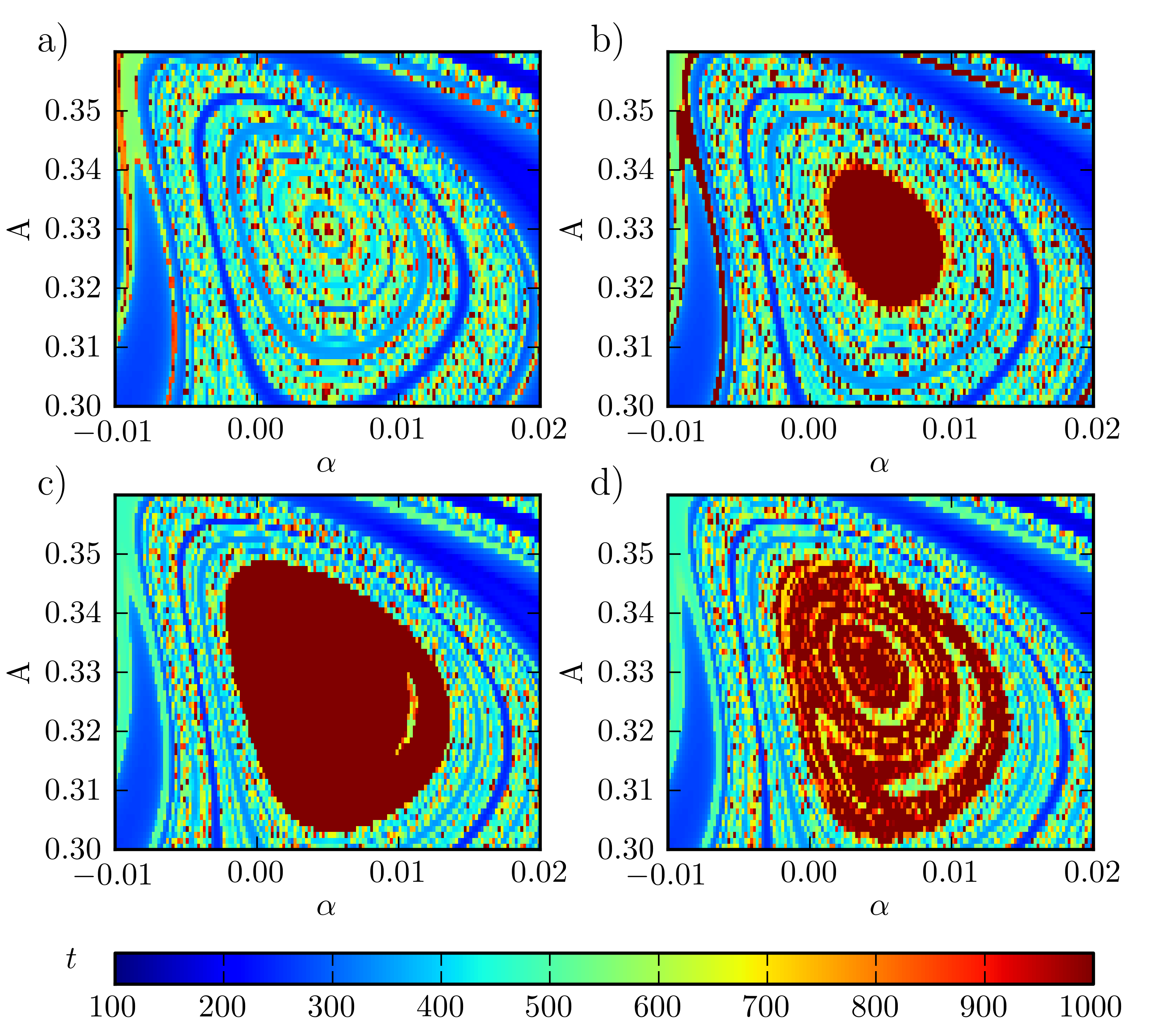}
  \caption{\label{fig:GrowingSaddle}
  (color online)
    Growth of the turbulent saddle via the emergence of a chaotic
    attractor followed by its  destruction in a boundary crisis.
    Visualized is a small region of
    the saddle in a projection similar to \reffig{fig:SaddleIntro} but
    spanned by the new pair of periodic orbits at $Re=250$ (with the phase such that energy is maximal.)
    Within the pre-existing chaotic saddle at
    $Re=248.5$ (a) the saddle-node bifurcation creates the new pair of two periodic
    orbits, of which one is stable. Its basin of attraction at  $Re=249.1$ (b) just after
    the bifurcation is visible as the red region of initial conditions
    that never reach the laminar state. On further increasing Re, the
    orbit undergoes a bifurcation cascade generating a chaotic
    attractor, shown at $Re=250.1$ (c), whose basin of attraction has
    grown. At $Re_c=250.13$ the attractor collides with its boundary in a crisis bifurcation.
    Beyond the crisis bifurcation, the attractor has turned into
    a repellor and it is now a part of the turbulent saddle. The remnant of the
    attractor's basin of attraction remains visible at  $Re=250.25$ (d), where now
    the new ``inner'' saddle is immersed in the ``outer'' one.
  }
\end{figure}

How does the complexity of the saddle increase? We explore one
  mechanism and specifically demonstrate a sequence of bifurcation
events, in which an infinite number of invariant
solutions is added to the saddle, leading to an increase in lifetimes; the mechanism bears
a strong resemblance with the events at the creation of the saddle as summarized above.
At $Re=249.01$ a local saddle-node bifurcation
creates a pair of new solutions which are not fixed points as the
  NBC-states but periodic orbits of period $T\simeq 60$. Of those,
as for the NBC-states, the lower branch has a single unstable direction and the upper branch solution is an attractor.
The stable periodic orbit is surrounded by its basin of attraction,
visible in \reffig{fig:GrowingSaddle}b) at $Re=249.1$.
As $Re$ increases, the orbit
undergoes a bifurcation cascade, creating infinitely many invariant states and
generating a local chaotic attractor (panel c).
On further increasing Re, the
chaotic attractor grows until at $Re_c=250.13$ it collides with its own
basin boundary in a crisis bifurcation. While the signature of the basin of attraction of
the former attractor is still visible for $Re > Re_c$ (panel d),
trajectories may now escape from the newly created ``inner'' saddle into the already existing
``outer'' one. Thus at the crisis bifurcation the infinitely many invariant solutions from the former
chaotic attractor are added to the turbulent saddle.
As before, the two-dimensional visualization captures the essential features of the 
process since the lower-branch orbit has only a single unstable direction and hence
a stable manifold of co-dimension one.

\begin{figure}
  \includegraphics[width=\linewidth]{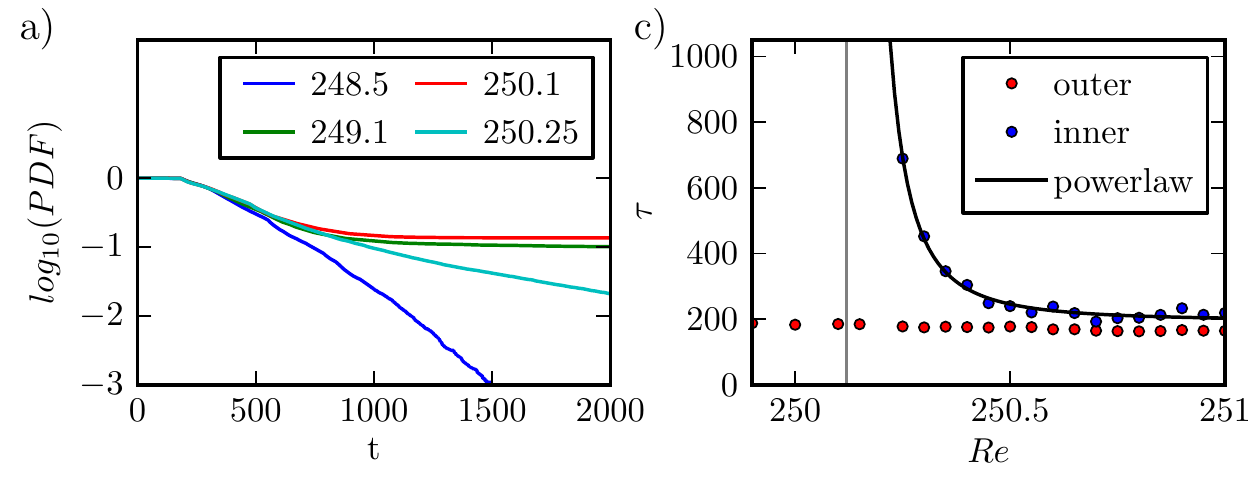}\\
  \includegraphics[width=.8\linewidth]{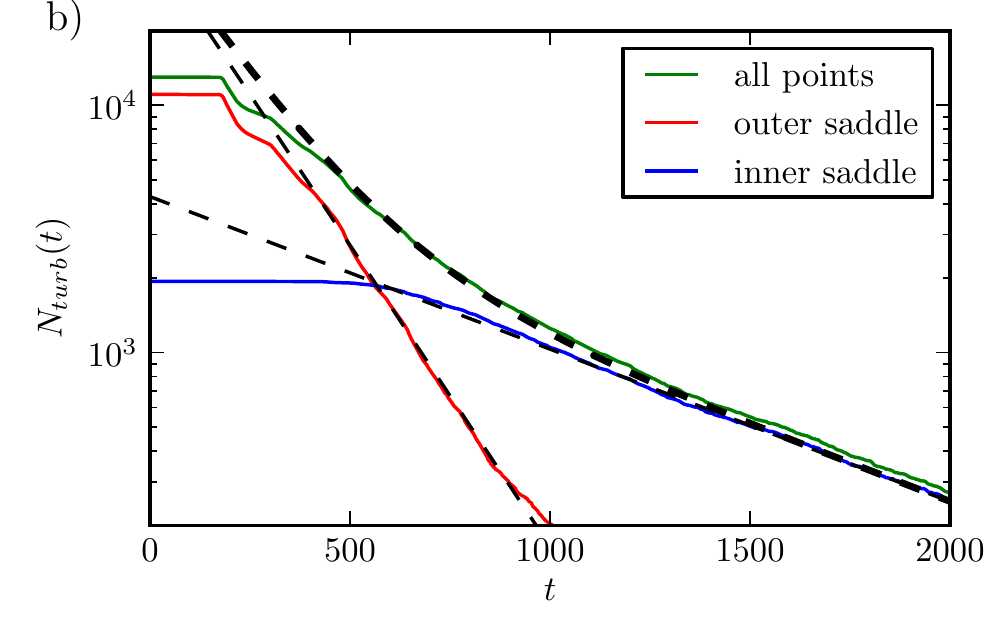}
  \caption{\label{fig:GrowingLifetime}
    (color online) Variations of lifetimes with Reynolds number. (a) 
    Survival probability calculated from the data in
    \reffig{fig:GrowingSaddle} for all four Re.
    Before the saddle-node bifurcation ($Re=248.5$), survival probabilities are
    exponentially distributed. After the bifurcation, the 
    attractor causes an offset corresponding to the initial conditions
    that never decay.
    After the boundary crisis (Re=250.25), the distribution is non-exponential.
    b) Detailed analysis of the lifetime statistics at $Re=250.25$, with
    the total number of trajectories that have not decayed up to a given time t as a function of time.
    In \reffig{fig:GrowingSaddle}d the shape of the former attractor is still visible and allows to calculate the characteristic
    lifetime separately for initial conditions from the new, inner saddle and the outer saddle.
    Both show exponential tails with a characteristic lifetime of $700$ and $180$, respectively.
    The thick dashed line on top shows the weighted average of the two exponentials -- it perfectly fits the actual 
    distribution apart from the trajectories that decay immediately.
    c) Reynolds number vs. characteristic lifetime $\tau$ as a
    function of $Re$, calculated separately for initial conditions from the inner
    and outer saddle. 
    Lifetimes from the outer saddle are almost constant.
    For the data from the inner saddle, it is infinite up to
    the point of the crisis bifurcation, after which it drops drastically with a power-law behavior and levels 
    slightly above the value of the outer saddle, showing that lifetimes have increased.
  }
\end{figure}

Over the sequence of these bifurcations,
the decay rate from the turbulent saddle to the laminar state varies non-monotonically.
Before the saddle-node bifurcation at $Re = 249.01$,
the survival probabilities are distributed exponentially (\reffig{fig:GrowingLifetime}a).
Beyond $Re = 249.01$, there are
initial conditions that
are trapped inside the stable region and never decay. Thus, the overall distribution is
a linear superposition of an exponential and a constant.
The contribution of the non-decaying trajectories 
increases with higher Reynolds numbers as
the attractor grows to fill more of the projection plane.
After the boundary crisis at $Re=250.13$,  the new bubble becomes a part of
the turbulent saddle.
Initially the escape
rate from the newly created embedded part of the saddle differs
substantially from the outer saddle's escape rate. Thus, for short
observation times the distribution is a linear superposition of two
different exponentials resulting in a non-exponential variation.

To quantify the variation of lifetimes due to both parts of the saddle,
we separate initial conditions into two groups corresponding to the
location of the former attractor in the discussed projection plane and
compute lifetimes for both parts of the saddle independently.
This is detailed for $Re=250.25$ in \reffig{fig:GrowingLifetime}c, where instead
of a \emph{normalized} survival probability we plot
the absolute number of initial conditions that survive for a given
time.
Since the initial conditions are uniformly distributed over the area shown
in \reffig{fig:GrowingSaddle} the absolute numbers are a measure of the state
space area that contributes to a given lifetime.
The total number of data points is separated into those
initial conditions from the inner and outer saddle.
In \reffig{fig:GrowingLifetime} both individual distributions show clear exponential tails
with characteristic lifetimes of $700$ and $180$, respectively.
The sum of the two exponentials, weighted by the number of initial conditions in the two regions,
(thick black line) almost perfectly recovers the total
distribution. Similar non-exponential distributions
  due to several weakly connected saddle structures have also been
  studied in Hamiltonian systems, where nested saddles are shown to
  generate algebraic lifetime distributions \cite{Hanson1985}.

The variation of $\tau$ with Re
for both the inner and outer saddle region is
shown in \reffig{fig:GrowingLifetime}(b).
While it remains almost constant for the outer region, in the inner region
the lifetime diverges
while the attractor exists and drops after the boundary crisis, 
well approximated by a power law with exponent $-2.1$.
The characteristic lifetime levels off slightly above the value of the outer saddle, indicating that trajectories
that leave the inner saddle have to first pass through the outer
one before decaying.
Consequently, those trajectories have longer lifetimes.
At higher Re, additional heteroclinic bifurcations will enhance
the dynamical connectivity between the inner and outer saddle region
which finally translates the higher lifetimes from the inner region to the
whole saddle. Thereby, the characteristic lifetime of the saddle
increases when it expands due to the incorporation of the newly created turbulent bubble.

We have found sudden expansions of the structure of the chaotic saddle at discrete
Reynolds numbers as discussed for $Re\simeq 250$  at various Re,
 indicating that the mechanism is a generic way to increase the complexity of the
state space.
While the presented example of a \emph{stable} periodic orbit emerging in a
saddle-node bifurcation followed by a chaotic cascade and a boundary
crisis of a transiently existing attractor is observed for many parameters in our system,
a more general variant of the discussed mechanism starts with \emph{unstable} states.
The sequence of events described above then appears in a weakly unstable subspace of
the full system.
This does not involve a global attractor for intermediate Re and is hence more
difficult to detect as the lifetimes do not diverge.
Yet, the same scenario still increases the complexity of the
saddle, leading to growing lifetimes.

The formation of turbulent bubbles and their destruction at discrete Re values
constitutes abrupt changes in the state space of shear flows.
The described mechanism naturally leads to non-smooth and non-monotonic
variations of $\tau$ as well as non-exponential lifetime statistics.
While we have found these features
for intermediate Re in the considered system they have not been
reported yet for
experimental lifetime studies. 
Apart from the symmetry restriction, which is only possible in numerics,
one challenge is that the bifurcations become dense in Re and begin to overlap.
To resolve this, very fine steps in Re are needed, as well as a very exact control
of the applied perturbation. 
In a sufficiently coarse grained analysis, an
apparent smooth variation emerges due to the collective
effect of many bifurcations. 

The mechanism we explored in this letter allows to
directly relate state-space structures and their variation to
statistical properties of transitional flows. While our
analysis is focused on plane Couette flow, recent investigations of
pipe flow and even mangnetohydrodynamic systems suggest that the described
mechanism by which turbulent saddles grow in
complexity are very generic. This emphasizes
the significance of low-dimensional dynamical systems theory for understanding the
physics of fluid mechanical systems with a subcritical transition to turbulence.

%

\end{document}